\documentclass[12pt]{article}
\usepackage{amssymb}
\begin{document}
\newcommand{\D}{\mathrm{d}}
\newcommand{\E}{\mathrm{e}}
\newcommand{\I}{\mathrm{i}}
\begin{titlepage}
\renewcommand{\thefootnote}{\fnsymbol{footnote}}
\vspace*{3.5cm}
\begin{center} \LARGE
{\bf  QCD$_{1+1}$ with Static Quarks as
Supersymmetric Quantum Mechanics}
\end{center}
\vspace*{1.0cm}
\begin{center}

{\bf M. Seeger\footnote{\tt mseeger@theorie3.physik.uni-erlangen.de}
and M. Thies\footnote{\tt thies@theorie3.physik.uni-erlangen.de}}

\vspace*{0.2cm}
{\em Department of Physics, State University of New York,
Stony Brook, NY 11794\\
\vspace{0.1cm}
and
\vspace{0.1cm}
\\
Institute for Theoretical Physics, University of
Erlangen--N\"{u}rnberg,
\\
Staudtstr.\ 7, 91058 Erlangen, Germany\/\footnote{Home institution of both
authors}}
\vspace*{0.2cm}

\end{center}

\vspace*{2.0cm}

\begin{abstract}

We reexamine the solvable model problem of two static, fundamental
quarks interacting with a SU(2) Yang--Mills field on a spatial
circle, introduced by Engelhardt and Schreiber. If the quarks are
at the same point, the model exhibits a quantum mechanical
supersymmetry. At finite separation, the supersymmetry is explicitly
broken in a way which naturally explains the geometrical nature of
spectrum and state vectors of this system.

\end{abstract}

\vskip 4.0cm

\end{titlepage}

\setcounter{page}{2}

\addtocounter{footnote}{0}

\newpage

\setcounter{equation}{0}

The problem of two opposite static electric charges in QED$_{1+1}$
on a spatial circle
is a rather trivial one: In a canonical formulation, the only gauge invariant
degree of freedom of the Maxwell field is the zero mode of $A_1(x)$.
Its quantum mechanics is that of a free particle, totally decoupled
from the sources which only contribute a
c-number to the Hamiltonian, their (periodic)
Coulomb potential. By contrast, the analogue problem in Yang--Mills
theory on a circle exhibits a much richer dynamics. This was exposed in
detail in a recent gauge fixed canonical study \cite{ES95},
and further analyzed
from the point of view of Hamiltonian lattice gauge theory in
\cite{STY95}.

The non-Abelian SU(2) case shares with QED the property that only a single
quantum mechanical degree of freedom survives gauge fixing, here
the eigenphase of the  Wilson loop winding around the circle,
$\mbox{P} \exp \I g \int_0^L A_1(x)\D x$. However, as a compact variable
with the character
of a polar angle, this phase has
very different properties from the corresponding cartesian variable in the
Maxwell case. Moreover, it remains coupled
to the color spins of the static charges. As a result, an intricate
discrete
spectrum emerges which displays several remarkable features: If the
separation between the static quarks goes to zero, all but one states become
doubly degenerate. For finite distance $d$ between the sources, the energy
is strictly
linear in $d$, no matter how large the ratio $d/L$ is (this is not
the case in QED on a circle, since the periodic Coulomb potential is
a succession of inverted parabolae). Although this model has been
solved analytically in Ref.~\cite{ES95}, these features have remained
somewhat mysterious.

It is the purpose of this short note to point out that an
underlying quantum mechanical supersymmetry is responsible for these
regularities.
In this way, we hope to round off the discussion of this
instructive toy model and enhance its pedagogical value. More importantly,
 we
find it interesting that a model which was not manufactured to be
supersymmetric in the first place can develop such a symmetry
dynamically.
In view of the fact that we are dealing with
an (admittedly, extremely simple) non-Abelian gauge theory, we feel
that this phenomenon deserves some attention.

We do not discuss the resolution of the Gauss law and the resulting
reduction of
the Hamiltonian to a quantum mechanical problem, as this has been
described elsewhere in great detail \cite{LTYL91,LNT94}. Instead, we
proceed immediately to the Hamiltonian given by Engelhardt and Schreiber
in Eq.~(14) of Ref.~\cite{ES95}.
It describes quantum mechanics of the variable $c$, proportional to
the eigenphase of the spatial
Wilson line, coupled to a two state system of the quark color spins.
If the quarks are sitting at the same point, $\mbox{Tr}\,Q_{\rm q}^2$
(with $Q_{\rm q}$ the quark color charge) is both gauge invariant and
conserved, therefore the natural basis vectors are the color singlet
and triplet states $|S\rangle$, $|T\rangle$. At finite separation between
the static sources, $\mbox{Tr}\,Q_{\rm q}^2$ ceases to be gauge invariant and
there is no particular reason to diagonalize it.
In Ref.~\cite{ES95},
a rotated basis $|\tilde{S}\rangle$, $|\tilde{T}\rangle$ has been
found to be most convenient which can be shown to diagonalize
the gauge invariant (but no longer conserved) operator obtained
from $\mbox{Tr}\,Q_{\rm q}^2$ by
inserting gauge strings between the two separated charges.

In order to clarify the structure of this Hamiltonian, we found it useful to
transform to ``radial''
wavefunctions
(i.e. pull out a Jacobian factor $1/\sin(\pi c)$), convert the
kinetic energy explicitly
to the  $\tilde{S}$, $\tilde{T}$ basis, and use the notation $\theta=\pi c$,
$d=y-x \geq 0$ ($\theta$ as a reminder that this angle runs only
from 0 to $\pi$).
Then we obtain
\begin{equation}
{\cal H}=\frac{g^2}{4} \left( L {\cal H}_0 + d {\cal H}_1 \right) \ ,
\label{A1}
\end{equation}
where we have exhibited the dependence on the coupling constant $g$
(dimension of a mass), the size $L$ of the system and the distance $d$
between the sources, and introduced
the dimensionless Hamiltonians
\begin{equation}
{\cal H}_0 = \left( \begin{array}{cc} -\frac{1}{2} \frac{\partial^2}
{\partial \theta^2} - \frac{1}{2} & 0 \\ 0 &
-\frac{1}{2} \frac{\partial^2}
{\partial \theta^2} - \frac{1}{2} + \frac{1}{\sin^2 \theta} \end{array}
\right)   \ ,
\label{A2}
\end{equation}
\begin{equation}
{\cal H}_1 = \left( \begin{array}{cc} \frac{3}{2} & -\I \left(
\frac{\partial}{\partial \theta} + \cot \theta \right) \\
\I \left( - \frac{\partial}{\partial \theta} + \cot \theta \right) &
- \frac{1}{2} \end{array} \right) \ .
\label{A3}
\end{equation}
${\cal H}_0$ is the part which survives if we let the distance between the
charges go to zero. In its present form (\ref{A2}), one easily recognizes the
structure characteristic of a
``supersymmetric quantum mechanical system'' (for recent reviews cf.\
\cite{CKS95} and the monograph \cite{J96}), namely
\begin{equation}
{\cal H}_0 = \left( \begin{array}{cc} Q^+ Q^- & 0 \\ 0 & Q^- Q^+ \end{array}
\right)
\label{A5}
\end{equation}
with
\begin{equation}
Q^{\pm} = - \frac{1}{\sqrt{2}}\left( \pm \frac{\partial}{\partial \theta} +
\cot \theta \right) \ .
\label{A4}
\end{equation}
${\cal H}_1$, the part accounting for the $d$-dependence of the spectrum,
breaks the supersymmetry, but its structure is also closely related to the
operators $Q^{\pm}$,
\begin{equation}
{\cal H}_1 = \left( \begin{array}{cc} \frac{3}{2} & \I \sqrt{2} Q^+ \\
-\I \sqrt{2} Q^- & -\frac{1}{2} \end{array} \right) \ .
\label{A6}
\end{equation}
In order to clarify both the supersymmetry of ${\cal H}_0$
and the nature of the breaking term ${\cal H}_1$, we introduce the
(non-hermitean)
``supercharge operator"
\begin{equation}
{\cal Q}=-\I \sqrt{2}Q^- \sigma^- \ ,
\label{A7}
\end{equation}
where $\sigma^- = \left( \sigma^1 - \I \sigma^2 \right)/{2}$ is a
Pauli matrix playing the part of the fermion
creation operator.
In terms of ${\cal Q}$, ${\cal Q}^{\dagger}$, we then have the superalgebra
\begin{equation}
{\cal H}_0 = \frac{1}{2} \left\{ {\cal Q}, {\cal  Q}^{\dagger} \right\} \ ,
\qquad \left\{{\cal Q},{\cal Q}\right\} = 0 \ ,
 \qquad \left[ {\cal H}_0, {\cal Q} \right] = 0       \ .
\label{A8}
\end{equation}
As usual in applications of supersymmetry to quantum mechanics, the
``bosonic" and ``fermionic" sectors of the theory should not be taken
literally, but can refer to any two-state system. Here, they correspond
to the quark color spin ``singlet" ($\tilde{S}$) and ``triplet" ($\tilde{T}$)
sectors, where we reemphasize that for finite separations these
can be defined in a gauge invariant way by inserting appropriate
strings~\cite{ES95}. The supercharge together with the ``fermion number
operator"
\begin{equation}
{\cal N}_{\rm F} = \sigma^- \sigma^+ = \left( \begin{array}{cc} 0 & 0 \\
0 & 1 \end{array} \right)
\label{A9}
\end{equation}
allows us to express ${\cal H}_1$ in a more concise form as well,
\begin{equation}
{\cal H}_1 = \frac{3}{2} -2 {\cal N}_{\rm F} + {\cal Q} + {\cal Q}^{\dagger}
 \ .
\label{A10}
\end{equation}
Since ${\cal H}_0$ conserves both supercharge and fermion number, we see
at once that
\begin{equation}
\left[ {\cal H}_0, {\cal H}_1 \right] = 0 \ ,
\label{A11}
\end{equation}
a rather useful property
which immediately tells us that the eigenvalues will be linear in $L$ and $d$.

Having exhibited the symmetry structure of the Hamiltonian, we turn
to its diagonalization which now does not require any particular skills:
${\cal H}_0$ coincides with
one of the most elementary examples of supersymmetric quantum
mechanics, the infinite square well and its superpartner, a
P\"{o}schl--Teller potential \cite{PT33}. The trivial solutions of the
square well potential,
\begin{equation}
\langle \theta | n \rangle = \sqrt{\frac{2}{\pi}} \sin ( n \theta )
\ , \quad n=1,2,\ldots\ , \qquad \epsilon_n = \frac{1}{2}\left( n^2-1
\right)
\label{A12}
\end{equation}
yield at once the (normalized) degenerate solutions
 of the $1/\sin^2 \theta$ potential \cite{CKS95},
\begin{equation}
| \tilde{n} \rangle = \frac{1}{\sqrt{\epsilon_n}} Q^- |n \rangle \ .
\label{A13}
\end{equation}
In the case at hand,
\begin{equation}
\langle \theta | \tilde{n} \rangle = \frac{1}{\sqrt{n^2-1}}\left(
\frac{\partial}{\partial \theta} -\cot \theta \right) \langle \theta |
n \rangle
=\sqrt{\frac{2}{\pi(n^2-1)}}\left( n \cos n \theta -\cot \theta \sin n \theta
 \right)
\ .
\label{A13a}
\end{equation}
Here, we label the states in the ``fermionic" sector
by $2,3,\ldots$ so that degenerate ones have the same
index. The state with $n=1$ is missing because $Q^-$ annihilates $|0\rangle$.

For $d=0$ we can identify the bosonic sector with
a pointlike {\em color neutral} object and
the fermionic sector with a
pointlike {\em adjoint} charge. Hence after subtracting the masses, there
is a degeneracy between excited
states of pure Yang--Mills theory on the circle and
states containing an additional adjoint charge.
While it is perhaps not surprising
that the energies are the same as both involve strings winding around the
circle, the result is non-trivial since the gluonic wavefunctions
are different. In particular, their number of nodes
differs by one.
The groundstate is at zero energy and
non-degenerate, as it must be for unbroken supersymmetry.
Note that we did not have to subtract any constant from the Hamiltonian
in order
to achieve this, since the $-\frac{1}{2}$ term in Eq.~(\ref{A2})
arises from the transformation to radial wavefunctions as an ``effective
potential" \cite{LNT94}. The full ground state wavefunction
 is a constant ($\sin \theta$
is canceled by the Jacobian factor), which is evidently annihilated by
the kinetic energy operator.

Let us now turn to the case $d > 0$ where the supersymmetry gets
explicitly broken, and diagonalize ${\cal H}_1$. Writing
down the eigenvalue equation for
${\cal H}_1$ in terms of a vector with components $u, v$ and
the eigenvalue $\eta$, and
eliminating one of the two components, say $v$, we get
\begin{equation}
Q^+ Q^- u = \frac{1}{2} \left( \eta + \frac{1}{2} \right)
\left( \eta - \frac{3}{2} \right) u \ , \qquad
\left( \eta + \frac{1}{2} \right) v =-\I \sqrt{2} Q^- u \ .
\label{A14}
\end{equation}
Inserting the known eigenvalues $\frac{n^2-1}{2}$ of $Q^{+}Q^{-}$
from above, this yields the two solutions
\begin{equation}
\eta = \frac{1}{2}\pm n \ .
\label{A16}
\end{equation}
The full result for the eigenvalues of $H$ is therefore
\begin{equation}
E_n^{\pm} = \frac{g^2}{4} \left[ L \left( \frac{n^2-1}{2} \right)
+ d \left( \frac{1}{2} \pm n \right) \right]  \ ,
\label{A17}
\end{equation}
where $E_1^-$ should be omitted. The corresponding eigenfunctions
are
\begin{equation}
\Psi_n^{\pm}( \theta)= \frac{1}{\sqrt{\pi n (n \pm 1)}}
\left( \begin{array}{c} (\pm n + 1) \sin n\theta \\
\I \left( \cot \theta \sin n \theta - n \cos n \theta  \right) \end{array}
\right) \ ,
\label{A17a}
\end{equation}
again discarding the $(n=1, -)$ case. (The normalization factor
can be obtained without evaluating an integral, along the same lines
which
lead to Eq.~(\ref{A13})).
These results agree with those obtained in Ref.~\cite{ES95}, where however
the wavefunctions were not explicitly given.
Note that the state vectors depend neither on $g$ nor on $L$ or $d$
but are purely geometrical. The dependence of the energy on these
constants only comes through the Hamiltonian, cf.\ Eq.~(\ref{A1}).

Instrumental for the strict linear $d$-dependence
is the fact that ${\cal H}_0$ and ${\cal H}_1$ commute
(first order degenerate perturbation theory is exact in this case).
The supersymmetry gets explicitly broken by the finite separation
of the static quarks. Nevertheless, this symmetry is useful for
understanding the spectrum, in much the same way as rotational
symmetry is useful for understanding
the spectrum of the hydrogen atom in a uniform magnetic field.
In this familiar case, the interaction is also linear in the generators
of the
symmetry group, thus yielding a simple, geometrical
level splitting pattern.

It is worth emphasizing that the supersymmetry has nothing
to do with the residual gauge symmetry after gauge fixing related
to topologically non-trivial, ``large" gauge
transformations. For pure Yang Mills theory on a circle, the
corresponding ``center symmetry" is the reflection symmetry
around the point $\theta=\pi/2$, in accordance with the homotopy
group $\pi_1\left[\mbox{SU}(2)/\mathbb{Z}_2\right]
= \mathbb{Z}_2$ (see e.g. \cite{LST95}).
Unlike dynamical quarks,
static (fundamental) sources do not destroy the center symmetry,
since they are insensitive to the spatial
boundary conditions and therefore cannot distinguish between
periodic, ``small" and antiperiodic, ``large"
gauge transformations.
The $\mathbb{Z}_2$ symmetry of the full
Hamiltonian (\ref{A1}) is simply
\begin{equation}
{\cal H}(\theta) = \sigma_3 {\cal H}(\pi -\theta) \sigma_3 \ .
\label{A18}
\end{equation}
Thus, we can classify the eigenstates according to their
$\mathbb{Z}_2$ parity,
\begin{equation}
\sigma_3 \Psi(\pi - \theta) = \pm \Psi(\theta) \ .
\label{A19}
\end{equation}
This explains why the ``bosonic" and ``fermionic" components of our
solutions (\ref{A17a}) have
definite (but opposite) parity under $\theta \to \pi - \theta$.
We also note that the supercharge~(\ref{A7}) is invariant under
the center symmetry, so that there is no conflict between the
two symmetries.

Summarizing, we have identified a quantum mechanical supersymmetric
structure in a model consisting of two static quarks and a SU(2)
Yang--Mills field on a circle. As demonstrated in previous works
\cite{ES95,STY95}, it is not necessary to account for this particular
symmetry in order to solve the model. However, it is then difficult
to understand
the dependence of the spectrum on the distance between the
quarks, in particular the degeneracies at $d=0$. The emergence of
a supersymmetry in a model where it has not been put in ``by hand"
is interesting in its own right. As an additional bonus, supersymmetry
helps to determine the full spectrum and set of eigenvectors of
this model with remarkably little effort.

It will be interesting
to see whether this line of attack is useful for dealing
with other types of non-Abelian gauge theories.

\vskip 1.0cm
We should like to thank the members of the Nuclear Theory Group at
Stony Brook, where
this work was carried out, for the hospitality extended to us.
Financial support from the Volkswagen-Stiftung is gratefully acknowledged.

\bibliographystyle{unsrt}

\end{document}